\documentstyle[epsf,aps,multicol,pre,tighten,amssymb]{revtex}

\begin{document} 
\draft 

\title{Statistics of finite-time Lyapunov exponents in the Ulam map}                
 
\author{Celia Anteneodo}

\address{Centro Brasileiro de Pesquisas F\'{\i}sicas,
         R. Dr. Xavier Sigaud 150, \\
         22290-180, Rio de Janeiro, Brazil\\
}
 
\date{\today} 
\maketitle 
 
\begin{abstract} 
The statistical properties of finite-time Lyapunov exponents at the Ulam 
point of the logistic map are investigated. 
The exact analytical expression for the autocorrelation function of 
one-step Lyapunov exponents is obtained, allowing the calculation of 
the variance of exponents computed over time 
intervals of length $n$. The variance anomalously decays as $1/n^2$.
The probability density of finite-time exponents noticeably deviates from 
the Gaussian shape, decaying with exponential tails and presenting  
$2^{n-1}$ spikes that narrow and accumulate close to the 
mean value with increasing $n$. 
The asymptotic expression for this probability distribution 
function is derived. 
It provides an adequate smooth approximation to describe numerical 
histograms built for not too small $n$, where the finiteness of 
bin size trimmes the sharp peaks. 
\end{abstract} 
 

 
\begin{multicols}{2} 

\narrowtext

\section*{Introduction}          

In a dynamical system, Lyapunov exponents (LEs) quantify the average 
exponential rate of expansion or contraction of infinitesimal volume 
elements in each direction of phase-space. 
Negative exponents imply convergence of initially nearby trajectories, 
while positive ones mean exponential divergence of neighboring orbits 
therefore signaling chaotic behavior. 

By the theorem of Oseledec \cite{oseledec,ruelle}, assuming ergodicity,  
each exponential rate of growth converges to a time asymptotic limit,  
$\lambda_\infty$, independently of the particular trajectory chosen 
(for almost all initial conditions).  
A useful generalization of asymptotic LEs are 
the local finite-time LEs, $\lambda_n$, calculated over a time 
interval of length $n$ along a given trajectory \cite{ott}. 
In contrast to asymptotic ones, exponents defined for finite time 
depend on the initial conditions. 
Local finite-time exponents must be distinguished from other local 
exponents also considered in the literature such as the so-called local 
finite-sample exponents \cite{localexp}. 
The analysis of finite-time LEs is particularly important as they are 
the quantities actually measured in computational studies, 
unavoidably performed for finite time. 
Local LEs have proved to be useful in the 
characterization of a variety of phenomena ranging from two-dimensional 
turbulence \cite{turbulence} to the so-called Loschmidt echo \cite{echo}. 
These local exponents are fluctuating quantities that may even change 
sign depending on the degree of heterogeneity of the relevant subset of 
phase space.
Their fluctuations give rise to a nontrivial probability density 
$P(\lambda_n)$ that asymptotically collapses to a Dirac delta 
function centered at $\lambda_\infty$.  
The deviations from this limit and the convergence to it contain 
rich information on the underlying dynamics \cite{ott,beck}.

A relevant question on the dynamics of chaotic systems 
is the existence or not of true trajectories lying 
close to the numerically generated ones. 
It is known that finite-time LEs fluctuating around zero cause 
the nonexistence of such shadowing trajectories \cite{silvina}. 
Particularly, in the synchronization transition of lattices of coupled 
chaotic one dimensional (1D) maps \cite{cml1}, 
unshadowability of chaotic orbits is due to 
unstable dimension variability, characterized by the second largest LE 
fluctuating about zero \cite{cml2}. 
The connection between the probability distribution function (PDF) of 
finite-time LEs in the synchronized states and the PDF of the local 
exponents in the uncoupled map \cite{preprint} motivates the study of the 
statistical properties of finite-time LEs in chaotic maps. 
Within this context, the numerical nature of the difficulties 
involved makes analytical results specially relevant.

This work focuses on the statistical properties of finite-time 
LEs for the logistic map at a parameter value yielding fully 
developed chaos, namely, at the Ulam point, where $x\mapsto 4x(1-x)$. 
This particular mapping allows exact calculations that are expected 
to be useful for more general systems belonging to the same class 
of chaotic behavior. 
We begin by calculating the autocorrelation 
function of one-step LEs. This function allows to evaluate exactly 
the variance of local LEs. 
Furthermore, we investigate thoroughly the PDF of finite-time LEs. 
Taking as a starting point the work by
Prasad and Ramaswamy \cite{pram}, we extend it by deriving 
an user-friendly expression for the PDF of finite-time LEs, 
a very good approximation for not too small $n$.

\section*{Finite-time Lyapunov exponents}

For a 1D map $x \mapsto f(x)$ with differentiable $f$, finite-time LEs     
$\lambda(n)$, computed over the interval of length $n$, 
are given by \cite{beck}

\begin{equation} 
\lambda_n \;=\; \frac{1}{n}\sum_{j=0}^{n-1}
\ln|f^\prime(x_j)| ,
\label{lambda} 
\end{equation} 
where $x_j=f^j(x_0)$ is the $j^{th}$ iterate of the initial condition $x_0$. 
We will compute local LEs after a transient has elapsed, so that averages 
over different realizations can be calculated 
with the weight given by the invariant density $p(x)$ of the chaotic attractor. 

Random initial conditions yield fluctuations in the values of $\lambda_n$  
around $\langle \lambda_n \rangle = \lambda_\infty$,  
with variance $\sigma^2(n)=\langle \lambda^2_n \rangle-\lambda_\infty^2$, that
by means of (\ref{lambda}) can be expressed as 

\begin{equation}  \label{sigma2}
\sigma^2(n) \;=\; \frac{1}{n^2}\sum_{j=0}^{n-1} \sum_{k=0}^{n-1} C_{j,k}, 
\end{equation}
where

\begin{equation} \label{lncorre}
C_{j,k}\;=\; \langle  \ln|f'(x_j)|  \;  \ln|f'(x_k)| \rangle
- \langle  \ln|f'(x_j)| \rangle \langle \ln|f'(x_k)| \rangle
\end{equation}
is the two-time autocorrelation function of one-step (unitary time) 
LEs \cite{fujisaka}.

Separating diagonal from off-diagonal terms, one gets
 
\begin{equation}  
\sigma^2(n) \;=\;
\frac{1}{n^2} \sum_{k=0}^{n-1} C_{k,k} \;+\; 
\frac{2}{n^2}\sum_{j=0}^{n-1} \sum_{k>j}^{n-1}   C_{j,k} .
\end{equation}
If stationarity can be attained, 
then temporal translational invariance holds. 
In such case, correlations depend on the 
two times only through their difference, so that 

\begin{equation} \label{trans}
 \sum_{j=0}^{n-1} \sum_{k>j}^{n-1}  C_{j,k}
\;=\; 
\sum_{m=1}^{n-1} [n-m]  C(m),
\end{equation} 
where $C(m)\equiv C_{0,m}$, 
and additionally $ C(0)\equiv C_{0,0}=C_{m,m}$, $\forall m$.
Hence, 

\begin{equation}  \label{sigma2f}
\sigma^2(n) \;=\;
\frac{1}{n}  C(0) \;+\; 
\frac{2}{n^2}\sum_{m=1}^{n-1} [n-m] C(m).
\end{equation}
Higher moments would require the calculation of correlation functions 
involving several times. 

In order to evaluate the variance explicitly, one must compute the 
autocorrelation function $C$ characteristic of 
the particular mapping $f$ considered. 
In what follows, we will perform the explicit calculations for  
the logistic map $f(x)=\mu x(1-x)$ at $\mu =4$, with $x\in [0,1]$ 
(usually called Ulam map).

\section*{Time correlations in the Ulam map}

For this logistic map the dynamics is exactly solvable 
(see, for instance, \cite{ott,lili}). 
In fact, using the change of variables $z=2\arcsin(\sqrt{x})/\pi$, 
it is easy to find that the $j^{th}$ iterate is $ x_j=\sin^2(2^{j-1} \pi z_0)$. 
The time evolution for the new variable is given by 
the {\em tent} map $z\mapsto 1-2|z-1/2|$, with $z\in[0,1]$\cite{ott,lili}. 
The invariant density, $p(x)=[\pi\sqrt{x(1-x)}]^{-1}$, becomes  
uniform for $z \in [0,1]$. 
In the $z$-representation, the required correlations can be straightforwardly 
obtained by following the procedure of Nagashima and Haken \cite{modulation}  
to calculate the time autocorrelation function $C_H(m)$,  

\begin{equation} \label{correl}
C_H(m) \,=\, \left\langle H\bigl(g^m(z)\bigr) H^*(z) \right\rangle - 
\left\langle H\bigl(g^m(z)\bigr) \right\rangle  
\left\langle H^*(z) \right\rangle ,
\end{equation}
of an arbitrary function $H(z)$ where $z$ evolves through the mapping $g(z)$ 
and $*$ denotes complex conjugate. The basic idea is to consider 
the Fourier expansion

\begin{equation} \label{fourier}
H(g^m(z))=\sum_{k=-\infty}^{\infty} a_k^m {\rm e}^{2\pi ikz}, \;\;\;\;\;
m=0,1,2,\ldots,
\end{equation}
where

\begin{equation}
a_k^m=\int_0^1 dz\,H(g^m(z)){\rm e}^{-2\pi ikz}.
\end{equation}
In our case $H(z)=\ln|4\cos(\pi z)|$, then it is indifferent if we evolve $z$ either 
with the tent map or with the Bernoulli shift $g(z)=2z ({\rm mod}\;1)$ that will 
generate the same sequence $H(g^m(z))$. 
We will consider the latter case in order to use the results of \cite{modulation}.  
Recursively, they arrive at

\begin{equation} \label{ak}
a_k^m =  
\left\{ \matrix{    a_j^0     & \mbox{if   $k=2^m j$; $j\in \cal{Z}$} \vspace*{4mm}\cr 
                     0       & \mbox{otherwise}. } \right.
\end{equation}
Then, using (\ref{correl}), (\ref{fourier}), (\ref{ak}) and additionally 
considering that $\langle H(z) \rangle =\langle H(g^m(z)) \rangle=a_0^0, \;\forall m$, 
the correlation results 

\begin{equation} \label{CHm}
C_H(m) \,=\, \sum_{k\neq 0} a_k^0\,a^{0\,*}_{2^m k}. 
\end{equation}

Now, for our particular case we have 

\begin{equation}
a_l^0= \frac{2}{\pi}\int_0^{\pi/2} dy\,\ln(\cos y) \cos(2ly) \, ,
\end{equation}
that gives

\begin{equation}
a_l^0 \;=\; 
\left\{ \matrix{ 
 -\ln 2        \;,\;\;\;  & \mbox{if $l=0$}  \vspace*{4mm} \cr
   -\frac{(-1)^l}{2l} \;,\;\;\;     & \mbox{otherwise}. } \right. 
\end{equation}
Substituting these coefficientes into (\ref{CHm}), 
we obtain the following exact expression 
for the stationary autocorrelation of one-step LEs: 

\begin{equation} \label{correlation}
C(m) \;=\; 
\left\{ \matrix{ 
\frac{\pi^2}{12}  \;,\;\;\; & \mbox{if $m=0$}  \vspace*{4mm} \cr
   -\frac{\pi^2}{24} \frac{1}{2^{m}}  \;,\;\;\;     & \mbox{otherwise}. } \right. 
\end{equation}
The autocorrelation function decays exponentially fast with time, 
with a characteristic time that is the inverse of $\lambda_\infty=\ln 2$.  
Although the sequence $\{x_j\}$ generated by the Ulam map is 
$\delta$-correlated (more precisely, 
$C_x(m)=\langle x_j x_{j+m}\rangle-\langle x_j\rangle\langle x_{j+m}\rangle=
\frac{1}{8}\delta_{0m}$ \cite{correl1}), correlations of higher powers 
of $x_j$ do not decay to zero immediately\cite{correl2}, 
leading to the exponential decay of the autocorrelation of one-step LEs. 

Once the autocorrelation (\ref{correlation}) is known, from Eq. (\ref{sigma2f}), 
we obtain

\begin{equation} \label{sigma_log}
\sigma^2(n) \;=\; \frac{\pi^2}{6\,n^2} \left( 1-\frac{1}{2^n} \right)   .
\end{equation}
This result valid for all $n$ is in agreement with the outcomes of 
numerical experiments\cite{pram,fujisaka,ts95} as well as with an 
asymptotic expression previously found \cite{ts95}. 
It is noteworthy that the variance decays with time as $1/n^2$. 
Then the quantity 
$\tilde\lambda_n=n(\lambda_n-\lambda_\infty)\;=\; \sum_{j=0}^{n-1}
\ln|f^\prime(x_j)/2|$ does not evolve diffusively, since its variance tends 
asymptotically to a constant value. 
If the terms of the sum in Eq. (\ref{lambda}) 
were independent, then the decay of $\sigma^2(n)$ should be as $1/n$. 
However, those terms are highly dependent: 
The sum can be expressed as a function of the single variable $x_0$ 
through the map successive iterates $f^j(x_0)$.  
This is a nice example illustrating that linear 
correlation and dependence are quite different statistical concepts.  
Naturally, although the correlations decay exponentially fast, 
the central limit theorem will not hold in this case.

\section*{PDF of finite-time exponents in the Ulam map}

Given a PDF $\rho_X(x)$, the standard formula for the PDF 
$\rho_Y(y)$, such that $y=\phi(x)$, is 

\begin{equation} \label{probformula}
\rho_Y(y)=\int \rho_X(x)\delta(y-\phi(x)).
\end{equation}

Prasad and Ramaswamy have already presented \cite{pram} 
the  exact although hermetic expression for the PDF  
$P(\lambda_n)$ of local LEs on the Ulam map, 
that comes immediately from the knowledge of the invariant measure $p(x)$, 
using (\ref{probformula}). 
In fact, from Eq.(\ref{lambda}), one can write  
$\lambda_n=\ln[G_n(x)]/n$, with $G_n(x)=4^n \prod_{j\geq 0}^{n-1} |1-2f^j(x)| $.
Then, one has

\begin{equation} \label{hermetic0}
P(\lambda_n) = 
\int_0^1 dx \;p(x)\;\delta(\lambda_n-\ln[G_n(x)]/n),
\end{equation}
giving

\begin{equation} \label{hermetic}
P(\lambda_n) = n \exp(n \lambda_n) \sum_{roots} \frac{p(x)}{|G_n'(x)|},
\end{equation}
where the summation runs over all real roots of the $(2^n-1)$-order polynomial 
$G_n(x)-\exp(n\lambda_n)$.
Because $G_n(x)$ is odd order, there is always at least one real root. 
For $\lambda_n$ below the mean value, all roots are real. As $\lambda_n$ increases 
above the mean value, the roots leave the real axis by pairs. At each point where 
this occurs, $G_n'(x)=0$, then a new divergence appears in the PDF  
\cite{pram}. 
The shape of the exact PDF for the scaled exponent  
$\tilde\lambda_n=n(\lambda_n-\lambda_\infty)$ is illustrated in Fig.~1  
for different values of $n$. 

Prasad and Ramaswamy proposed a smooth ansatz for the  
discontinuous expression (\ref{hermetic}), 
that in terms of the scaled variable reads  

\begin{equation} \label{approx}
P(\tilde\lambda_n) = \frac{1}{\pi}\frac{ \exp(-|\tilde\lambda_n|)} 
{\sqrt{1 - \exp(- 2| \tilde\lambda_n|)}}, 
\end{equation}
for $\tilde\lambda_n \in (-\infty,n\ln 2]$. 
They based their conjecture on the analysis for small $n$. 
Here we will show that it is possible to advance further,  
developing expression (\ref{hermetic}) and 
evaluating it explicitly, therefore, 
finding analytically a smooth expression valid for large enough $n$.

\vspace*{-0.5cm}
\begin{figure}[ht] 
\begin{center} 
 {\epsfxsize 10.0cm \epsfbox{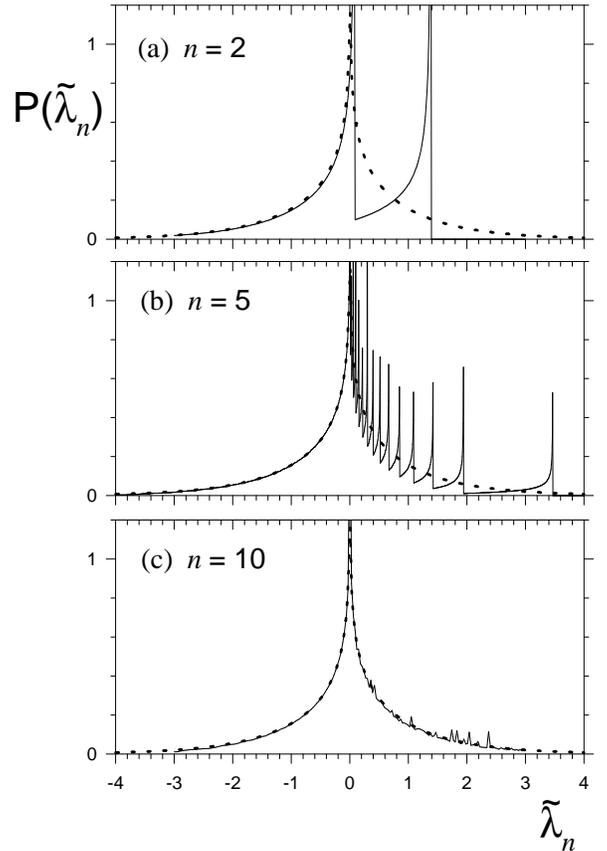}} 
\end{center} 
\vspace{-1cm} 
\caption{\protect Exact $P(\tilde\lambda_n)$ as a function of 
$\tilde\lambda_n\equiv n(\lambda_n-\lambda_\infty)$ (full lines), 
obtained numerically from Eq. (\ref{pn}), for 
different values of $n$ indicated in the figure. 
As $n$ increases divergences become narrower and cannot be 
appreciated unless one refines the mesh increasing the number of 
data exponentially with $n$. 
The dashed lines correspond to the smooth approximation given by 
(\ref{smooth}). 
} 
\label{fig1} 
\end{figure} 
\vspace{1cm}

Considering that $\sum_{j=0}^{n-1} \ln|f^\prime(x_j)| = \ln|[f^n(x_0)]^\prime|$ 
and making the transformation 
$x=\sin^2(\pi z/2)$, from Eq. (\ref{lambda}), one gets the compact expression 

\begin{equation} \label{fz}
\tilde\lambda_n  \;=\; -\ln[\sin(\pi z)] +\ln|\sin(2^n\pi z)|   \;\equiv F_n(z),
\end{equation}
where the scaled exponent is expressed as a function of the state variable 
$z$ (from now on the subindex 0 will be omitted)  evaluated at the instant 
at which the trajectory segment of time length $n$ starts. 
The first logarithmic term is a smooth function in $[0,1]$ that 
does not depend on $n$, while the 
second term is a periodic function with divergences at $z=j/2^n$, with $j\in{\cal Z}$ 
and $0\leq j \leq 2^n$. 
Increasing $n$ will increase the frequency and the comb of divergences will become 
more compact. 
The shape of $F_n(z)$ is illustrated in Fig.~2, for $n=5$. 

Since the invariant density of the attractor as a function of $z$ is uniform, 
using (\ref{probformula}), the PDF for $\tilde\lambda_n$ is given by

\begin{equation} \label{pn}
P(\tilde\lambda_n) = \int_0^1 dz \;\delta(\tilde\lambda_n-F_n(z)) = 
\sum_j \frac{1}{| F_n'(z^\star_j)|},
\end{equation}
where $z^\star_j$ are such that $F_n(z^\star_j)=\tilde\lambda_n$.

\vspace*{-0.5cm}
\begin{figure}[h] 
\unitlength 1mm 
\begin{center} 
 {\epsfxsize 8.7cm \epsfbox{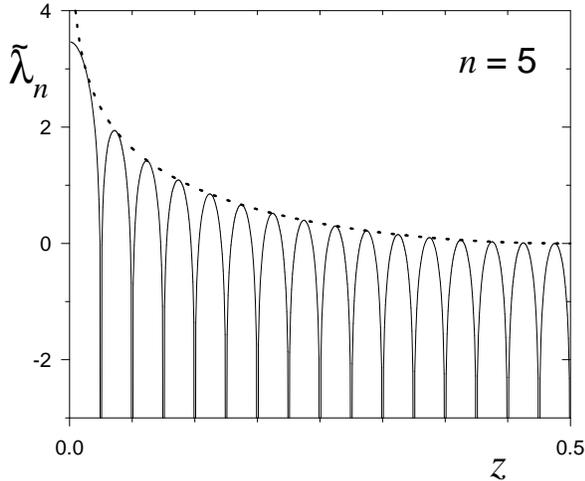}} 
\end{center} 
\vspace{-3cm} 
\caption{\protect 
Scaled Lyapunov exponent $\tilde\lambda_n \equiv F_n(z)$ as a function of $z$, 
the value of the state variable when the computation of the exponent is initiated 
(full lines). In this case $n=5$. 
The dotted line corresponds to the smooth component $-\ln[\sin(\pi z)]$ that 
does not depend on $n$. 
Increasing $n$ by one will double the number of periods. 
There is symmetry around $z=1/2$, then only half of the full abscissae interval has been 
represented.} 
\label{fig2} 
\end{figure} 
\vspace{1cm}

Let us separate the analysis into two cases according to the sign of $\tilde\lambda_n$. 

(I) For $\tilde\lambda_n<0$, the exact PDF is smooth. There is symmetry around $z=\frac{1}{2}$.  
In the interval $[0,\frac{1}{2}]$ there are $2^n-1$ 
intersections at $z^\star_{j\pm}=j/2^n\pm\delta_j$ such that $0<\delta_j<<j/2^n$. 
Taking into account that the main contribution to the derivative $F_n'$ 
comes from the periodic component, then 
$|F_n'(z^\star_{j\pm})|\simeq \pi 2^n |\cos(2^n\delta_j\pi)/\sin(2^n\delta_j \pi)|$, 
where, from (\ref{fz}),  
$ |\sin(2^n\delta_j\pi)| \simeq \exp(\tilde\lambda_n) |\sin(j\pi/2^n)|$.  
Substitution of the derivatives in (\ref{pn}) leads to

\begin{equation} \label{suma-}
P(\tilde\lambda_n) \simeq \frac{4 e^{\tilde\lambda_n}}{\pi 2^n} \sum_{j=1}^{2^{n-1}}  
\frac{|\sin(j \pi/2^n)|}{\sqrt{ 1 -e^{2\tilde\lambda_n}\sin^2(j \pi/2^n)| }}.
\end{equation}
Replacing the summation by an integral by identifying $x=j/2^n$, we get

\begin{equation} 
P(\tilde\lambda_n) \simeq \frac{4 e^{\tilde\lambda_n}}{\pi} 
\int_{0}^{1/2} dx \frac{\sin( \pi x)}{\sqrt{ 1- e^{2\tilde\lambda_n}\sin^2(\pi x) }} \,, 
\end{equation}
that gives 

\begin{equation} \label{smooth-}
P(\tilde\lambda_n) \simeq \frac{2}{\pi^2} 
\ln [\coth(-\tilde\lambda_n/2)] \, .
\end{equation}
The approximations here performed are expected to be good even for small $n$, namely 
$n \gtrsim 2$. In fact, a good agreement between the approximate and exact PDFs in this 
region is observed already for $n=2$ (see Fig.~1).

(II) The domain $0< \tilde\lambda_n \leq n\ln 2$ is more tricky. 
At each local maximum of $F_n(z)$, a divergence in the PDF appears, 
as it follows from (\ref{pn}). Then, in this region, 
the PDF possesses $2^{n-1}$ spikes. 
Let us consider the interspike intervals $(\tilde\lambda^-,\tilde\lambda^+]$ 
determined by two successive values of $z$ at which $F_n'=0$. 
In each interspike interval the PDF monotonously increases, 
because $|F_n'|$ decreases with increasing 
$\tilde\lambda_n$ (see Fig.~2), and diverges at the upper limit 
of the interval $\tilde\lambda^+$, where $F_n'=0$. 
These behaviors can be observed in Fig.~1, for $n=2$ and 5. 
As $n$ increases, although the cusps increase exponentially in number,
they get thinner, in such a way that they will become invisible 
in numerical histograms due to the finiteness of bin size. 
Then, if $n$ is large enough, histograms would look smooth although the 
exact PDF is not, unless one could make the bin size arbitrarily small. 
In such case, the higher the resolution, the larger the height of the spikes.  
However, in practice, there are constraints to reach an arbitrarily fine resolution
due to finiteness of numerical data sets. This fact turns it useful to find a 
smooth expression to describe the histograms actually observed 
for not too small $n$.  
In order to do so, we have to calculate the probability 
$\int_{bin} d\tilde\lambda P(\tilde\lambda)$ 
associated to a finite bin of the order of (or larger than) 
the interspike distance $\Delta\tilde\lambda=\tilde\lambda^+-\tilde\lambda^-$. 

For each diffeomorphic branch of $F_n$ one has 
$F_n'(z)=F_n'(F_n^{-1}(\tilde\lambda_n))=
1/[\frac{dF_n^{-1}(\tilde\lambda_n)}{d\tilde\lambda_n}]$, 
then, from Eq. (\ref{pn}), the probability that $\tilde\lambda_n$ belongs to the interval 
$(\tilde\lambda^-,\tilde\lambda^+]$ is given by

\begin{equation} \label{bin1}
\int_{\Delta\tilde\lambda} d\tilde\lambda P(\tilde\lambda)=
2\sum_{i=1}^m  \Bigl|  F_i^{-1}(\tilde\lambda^+) - F_i^{-1}(\tilde\lambda^-)  \Bigr|  
=2\sum_{i=1}^m \Delta z_i ,
\end{equation}
where $\Delta z_i$ are the segments displayed in Fig.~3, corresponding 
to the inverse images of $\Delta\tilde\lambda$. The summation 
runs up to a value $m$ that depends on the particular interval chosen 
and that is given by the smooth component of $F_n$.

\vspace*{-0.5cm}
\begin{figure}[h] 
\begin{center} 
 {\epsfxsize 8.5cm \epsfbox{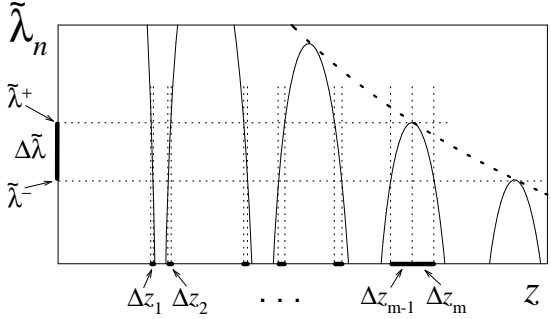}} 
\end{center} 
\vspace{-5cm} 
\caption{\protect 
Schematic representation of the scaled Lyapunov exponent $\tilde\lambda_n$ 
as a function of $z$. 
The segments on the horizontal axis represent the contributions to the probability 
associated to the interspike interval 
$\Delta \tilde\lambda \equiv \tilde\lambda^+  - \tilde\lambda^-$. 
} 
\label{fig3} 
\end{figure} 
\vspace{0.5cm}

This procedure will give rise to a discrete PDF where 
the integrated probabilities (\ref{bin1}) can be attributed, for instance, to 
the midpoints $\tilde\lambda^\star$ of each interval $\Delta\tilde\lambda$, such 
that

\begin{equation} \label{bin2}
P(\tilde\lambda^\star) \simeq 
\frac{ \int_{\Delta\tilde\lambda} d\tilde\lambda P(\tilde\lambda)}{\Delta\tilde\lambda}   
= 2\sum_{i=1}^m \frac{\Delta z_i}{\Delta\tilde\lambda} .
\end{equation}
Now, $\frac{\Delta\tilde\lambda}{\Delta z_i} \simeq |F_n'(z_i^\star)|$, 
where the derivative is evaluated at the midpoint $z_i^\star$ of the 
interval $\Delta z_i$. As the derivative $F_n'$ is dominated by the periodic component 
in Eq. (\ref{fz}), one can perform an analysis analogous to that for case (I), 
obtaining
\begin{equation} 
P(\tilde\lambda^\star) 
\simeq \frac{4 e^{\tilde\lambda^\star}}{\pi 2^n} \sum_{j=1}^{j_{max}}  
\frac{|\sin(j \pi/2^n)|}{\sqrt{ 1 -e^{2\tilde\lambda^\star}\sin^2(j \pi/2^n)| }},
\end{equation}
where $j_{max}\simeq  2^n\arcsin[\exp(-\tilde\lambda^\star)]/\pi$, for 
$\tilde\lambda^\star \in (\tilde\lambda^-,\tilde\lambda^+)$.
Replacing the summation by an integral as in case (I), one obtains 

\begin{equation} 
P(\tilde\lambda^\star) \simeq \frac{4 e^{\tilde\lambda^\star}}{\pi} 
\int_{0}^{\theta(\tilde\lambda^\star)} dx \frac{\sin( \pi x)}
{\sqrt{ 1- e^{2\tilde\lambda^\star}\sin^2(\pi x) }} \,,
\end{equation}
where $\theta(\tilde\lambda^\star)={\rm arcsin}[\exp(-\tilde\lambda^\star)]/\pi$.
After integration, one arrives at 

\begin{equation} \label{smooth+}
P(\tilde\lambda^\star) \simeq \frac{2}{\pi^2} 
\ln[\coth(\tilde\lambda^\star/2) ]\, .
\end{equation}
Let us recall that, in contrast to case (I), here we have 
a discrete PDF defined for the middle points of 
the interspike intervals. However, since the interspike distance 
$\Delta\tilde\lambda$ goes to zero 
for increasing $n$ and not too large $\tilde\lambda_n$, one can smooth the 
staircase function (\ref{smooth+}) by extending it to every point of the 
interval $(0,n\ln 2]$. In that case, one gets the symmetric counterpart 
of expression (\ref{smooth-}).

Expressions (\ref{smooth-}) and (\ref{smooth+}) can be gathered into the 
single expression 

\begin{equation} \label{smooth}
P(\tilde\lambda_n) \simeq \frac{2}{\pi^2} 
\ln( \coth|\tilde\lambda_n/2| )\, .
\end{equation}
This result explains the apparent symmetry of numerical PDFs 
for large $n$ despite the fact that left and righthand wings of the 
distribution have very different genealogies. 
For large $|\tilde\lambda_n|$, the PDF decays exponentially, 
yielding straight lines in the semi-log representation displayed in Fig.~4. 
Notice that the approximate distribution (\ref{smooth}) 
for the scaled variable $\tilde\lambda_n$  
does not depend on $n$. In particular, it is expected to be exact 
in the limit $n\to \infty$. That is, (\ref{smooth}) is a good  
representation of the delta function to which the finite-time 
distribution asymptotically collapses. 
Remarkably enough, the approximations that led to (\ref{smooth}) did not 
spoil the normalization condition, i.e., 
$\int_{-\infty}^\infty d\tilde\lambda P(\tilde\lambda)=1$ holds.

In Fig.~4 the approximation (\ref{smooth}) is compared to an 
histogram obtained from numerical iterations of the mapping.  
The analytical prediction is in very good accord with numerical results and 
it describes the numerical data more accurately than the 
ansatz (\ref{approx}).  
Notably, the variance of the scaled exponent 
$\tilde\lambda_n=n(\lambda_n-\lambda_\infty)$ calculated with our PDF  
is $\tilde\sigma^2=\pi^2/6$. Then for exponent $\lambda_n$ one obtains   
$\sigma^2(n)=\pi^2/(6n^2)$ which coincides with the exact 
result (\ref{sigma_log}), if neglecting the term of order $1/2^{n}$ 
(the order of the performed approximations). 
In contrast, the ansatz of ref. \cite{pram} gives 
$\sigma^2(n)=(\pi^2/12+[\ln 2]^2)/n^2 \simeq 0.79\,\pi^2/(6n^2)$, i.e., 
more than $20\%$ smaller than the true value.   

As an aside comment, notice that in the exact distribution the ordinate at 
the origin increases with $n$ but is finite for finite $n$. However, 
expression (\ref{smooth}) diverges at $\tilde\lambda_n=0$ 
(as also (\ref{approx}) does).  
In the exact PDF, divergences appear for positive 
abscissae only and accumulate close to the origin as $n$ increases. 
Therefore, strictly speaking, a divergence at the origin is expected 
in the limit $n\to \infty$ only.

\vspace*{-0.5cm}
\begin{figure}[htb] 
\begin{center} 
 {\epsfxsize 9.0cm \epsfbox{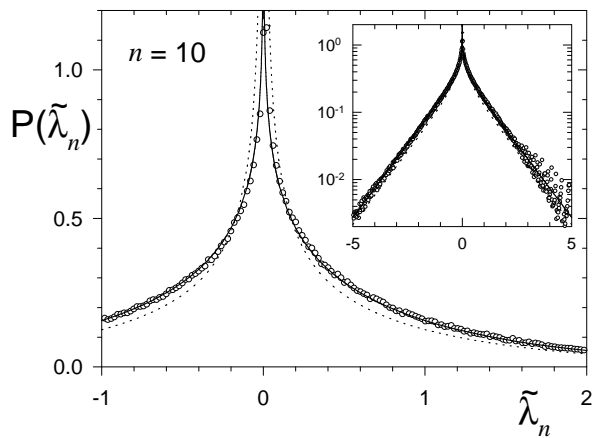}} 
\end{center} 
\vspace{-5cm} 
\caption{\protect $P(\tilde\lambda_n)$ as a function of $\tilde\lambda_n$ for $n=10$. 
Small symbols correspond to the histogram accumulated over $10^6$ histories 
obtained through numerical iterations of the map, after a transient of $10^3$ 
time steps. The bin size is 0.02.
Expressions (\ref{smooth}) (full line) and (\ref{approx}) (dotted line) 
are plotted for comparison. 
Inset: the same data in semi-log representation allow to appreciate the tails.
The spikes in the righthand tail of the histogram 
are visible for large $\tilde\lambda_n$ where they become more sparse and 
are amplified by the logarithmic scale.
} 
\label{fig4} 
\end{figure} 
\vspace{0.5cm}

\section*{Summary and final remarks}

We have derived analytical results for the statistics of finite-size exponents 
on the logistic map at the Ulam point. 
In particular, we have found the autocorrelation function of one-step LEs, 
which allows to obtain the exact analytical expression (\ref{sigma_log}) for 
the variance of local exponents, putting into evidence the origin of  
its anomalous time decay. 
Moreover, we have found the smooth analytical PDF (\ref{smooth}),   
expected to be the exact one in the limit $n\to \infty$.  
Therefore, we have extended the work initiated 
by Prasad and Ramaswamy\cite{pram}, 
providing a complete and consistent picture of the statistical 
properties of finite-time LEs at an outer crisis.

As the special shape of the distribution of finite-time exponents  
at the Ulam point is also seen in other systems presenting 
fully developed chaos \cite{pram}, our results are expected to be useful 
for more general systems falling into the same class.
In addition, these analytical results could give insights on features occurring in 
the synchronization transition of lattices of coupled chaotic maps \cite{cml1}, 
such as hiperbolicity breakdown through unstable dimension variability \cite{cml2}, 
since the distribution of finite-time LEs in the synchronized states of 
such networks is related to the distribution of 
local exponents 
in the uncoupled map \cite{preprint}. 

I am grateful to Sandro E. de S. Pinto and 
Ricardo L. Viana for very stimulating discussions,  
to Ra\'ul O. Vallejos for his usual insightful remarks, and
to Leonard A. Smith  and James Theiler for interesting comments. 
I acknowledge Brazilian agencies FAPERJ and PRONEX for financial support.

\end{multicols}


\begin{thebibliography}{99} 

\bibitem{oseledec} V.I. Oseledec, Trans. Mosc. Math. Soc. {\bf 19}, 197 (1968).

\bibitem{ruelle}    J.-P Eckmann and D. Ruelle, Rev. Mod. Phys. {\bf 57}, 617 (1985).

\bibitem{ott} E. Ott., {\em  Chaos in Dynamical Systems} (Cambridge University Press, 1993).

\bibitem{localexp} C. Ziehmann, L.A. Smith and J. Kurths, Phys. Lett. A {\bf 271}, 
237 (2000);  J.C. Vallejo, J. Aguirre and M.A.F. Sanju\'an, Phys. Lett. A {\bf 311}, 
26 (2003). 

\bibitem{turbulence} K. Nam, E. Ott, T. M. Antonsen, Jr., and P. N. Guzdar, 
Phys. Rev. Lett. {\bf 84}, 5134 (2000);
G. Boffetta, A. Celani, and S. Musacchio, Phys. Rev. Lett. 91, 034501 (2003). 

\bibitem{echo} F.M. Cucchietti, C.H. Lewenkopf, E.R. Mucciolo, H.M. Pastawski 
and R.O. Vallejos, Phys. Rev. E {\bf 65}, 046209 (2002).

\bibitem{beck} C. Beck and F. Schl\"ogl, {\em Thermodynamics of chaotic systems} 
(Cambridge University Press, 1993).

\bibitem{silvina}   S. Dawson, C. Grebogi, T. Sauer and J.A. Yorke, Phys. Rev. Lett. 
{\bf 73}, 1927 (1994). 

\bibitem{cml1}  A.M. Batista and  R.L. Viana, Phys. Lett. A {\bf 286},  134  (2001);
 A.M. Batista, S.E.D. Pinto, R.L. Viana and S.R. Lopes, Phys. Rev. E {\bf 65}, 056209 (2002).
 
\bibitem{cml2}  R.L. Viana, C. Grebogi, S.E. de S. Pinto, S.R. Lopes, A.M. Batista and 
J. Kurths, e-print nlin.CD/0310004, to appear in Phys. Rev. E. 

\bibitem{preprint} C. Anteneodo, S.E. de S. Pinto, A.M. Batista 
and R.L. Viana, Phys. Rev. E {\bf 68}, 045202(R) (2003).

\bibitem{pram} A. Prasad and R. Ramaswamy, Phys.  Rev. E {\bf 60}, 2761 (1999).

\bibitem{fujisaka}  H. Fujisaka, Prog. Theor. Phys. {\bf 70}, 1264 (1983).

\bibitem{ts95} J. Theiler and L.A. Smith, Phys. Rev. E {\bf 51}, 3738 (1995). 

\bibitem{lili} A.J. Lichtenberg and M.A. Lieberman, 
{\em  Regular and Chaotic Dynamics} (John Wiley \& Sons, 1992).

\bibitem{modulation} T. Nagashima and H. Haken, Phys. Lett. A {\bf 96}, 385 (1983).

\bibitem{correl1} S. Grossmann and S. Thomae, Z. Naturforsch. {\bf 32a}, 1353 (1977).

\bibitem{correl2} H.-P. Herzel and W. Ebeling, Phys. Lett. A  {\bf 111}, 1 (1985); 
errata: {\bf 111}, 457 (1985).


\end{thebibliography}
\end{document}